\documentclass[aps,prl,twocolumn,preprintnumbers,floatfix,reprint]{revtex4-1}
\usepackage{mathrsfs,natbib}
\usepackage{amsmath,amssymb, amsfonts, bbm}
\usepackage{graphics,epsfig,color,subfigure}
\usepackage{hyperref}
\usepackage{verbatim}
\usepackage{eufrak}

\newcommand{\be}{\begin{equation}}
\newcommand{\ee}{\end{equation}}

\def\bea{\begin{align}}
\def\ena{\end{align}}

\def\Tr{\mbox{Tr }}
\def\beqa{\begin{eqnarray}}
\def\enqa{\end{eqnarray}}

\hypersetup{
    colorlinks=true,       % false: boxed links; true: colored links
    linkcolor=red,          % color of internal links (change box color with linkbordercolor)
    citecolor=blue,        % color of links to bibliography
    filecolor=magenta,      % color of file links
    urlcolor=blue           % color of external links
}

\begin{document}

\title{Large N Volume Independence and an Emergent Fermionic Symmetry}

\author{G\"{o}k\c{c}e Ba\c{s}ar$^1$}
\email{basar@tonic.physics.sunysb.edu} 
\author{Aleksey Cherman$^2$}
\email{cherman@physics.umn.edu} 
\author{Daniele Dorigoni$^3$}
\email{d.dorigoni@damtp.cam.ac.uk} 
\author{Mithat \"{U}nsal$^4$}
\email{unsal.mithat@gmail.com} 
\affiliation{
$^1$ Department of Physics and Astronomy, Stony Brook University, Stony Brook, NY 11794, USA\\
$^2$ Fine Theoretical Physics Institute, Department of Physics, University of Minnesota, USA\\
$^3$ DAMTP, University of Cambridge, Wilberforce Road, Cambridge CB3 0WA, UK \\
$^4$ Department of Physics and Astronomy, SFSU, San Francisco, CA 94132 USA}

\preprint{FTPI-MINN-13/20, UMN-TH-3210/13}
\preprint{DAMTP-2013-31}

\begin{abstract}

Large-$N$ volume independence in circle-compactified QCD with  %$N_f \geq 1$
 adjoint Weyl fermions implies the absence of any phase transitions  as the radius is dialed  to arbitrarily small values.
This class of theories  are  believed to possess  a  Hagedorn density of hadronic states. It turns out that these properties are in apparent tension with each other, because a Hagedorn density of states typically implies a phase transition at some finite radius.  This tension is resolved if there are degeneracies between the spectra of bosonic and fermionic states, as happens in the  $N_f=1$ supersymmetric case.  Resolution of the tension for $N_f>1$ then suggests the emergence of a  fermionic symmetry at large $N$, where there is no supersymmetry.   We can escape the Coleman-Mandula theorem since the $N=\infty$ theory is free, with a trivial S-matrix.   We show an example of such a spectral degeneracy in a non-supersymmetric toy example which has a Hagedorn spectrum.
\end{abstract}

\maketitle
{\it Introduction.} Understanding the low energy properties of quantum chromodynamics (QCD) analytically is famously difficult due to strong coupling problems, even in the large $N$ limit. For some questions, the situation is better in supersymmetric (SUSY) theories, since fermionic symmetries enable powerful non-perturbative analytical techniques. However, the absence of SUSY in real world QCD prevents us from borrowing these techniques to explain its behavior.  In this paper, motivated by QCD, and for the reasons we detail below, we study asymptotically free QCD-like $SU(N)$ gauge theories with $N_f \ge 1$ massless Weyl fermions in the adjoint representation [QCD(Adj)], which have a quantitative connection to QCD at large $N$. We suggest that surprisingly, if two widely-believed properties of QCD(Adj) are indeed valid, for $N_f>1$ these theories should possess an emergent non-SUSY fermionic symmetry in the large $N$ limit. Apart from the QCD example discussed here, similar arguments are likely to apply to a much broader class of quantum field theories, for instance to fermionic extensions of two-dimensional principal chiral and Wess-Zumino-Witten models, which are relevant to condensed matter physics.

%%%%%%%%%%%%%%%%%%%%%
{\it The puzzle:}
%%%%%%%%%%%%%%%%%%%%% 
It has recently been understood that QCD(Adj) has the property of volume independence\cite{Kovtun:2007py} in the large-$N$ limit when compactified on $\mathbb{R}^3\times S^1$.  This is a working realization of the 
old Eguchi-Kawai proposal\cite{Eguchi:1982nm}.   
  This property does not hold for thermal compactification, where there is a phase transition to a deconfined phase at 
$  \beta_d= 1/T_d\sim \Lambda^{-1}$, the  strong scale.
  Volume independence applies to circle (spatial) compactifications, where fermions have periodic boundary conditions (PBCs), and the theory stays in the confined phase for arbitrarily small $L \sim N^0$. 
  
  To clarify the statement of volume independence and the distinction between spatial and thermal compactifications,  
let $H$  and  ${\cal H}$ denote the Hamiltonian  and the  Hilbert space of  the theory. 
Using  fermion number modulo two,  
$(-1)^F$,  we can split  ${\cal H}$  into   the bosonic   and fermionic    Hilbert spaces: 
 ${\cal H} \equiv  \cal B  \oplus  {\cal F}$.  
  The Euclidean path integral of the spatially-compactified theory computes the {\it twisted partition function}, 
\begin{align}
\label{eq:tpf}
\tilde{Z}(L) &\equiv \Tr (-1)^F e^{-L H} = Z_{\cal B} - Z_{\cal F} \nonumber \\
&= \int dM  [ \rho_{\cal B} (M) - \rho_{\cal F} (M) ] e^{-LM}
\end{align}
which is a graded trace, in contrast to the {\it thermal partition function}, which is
\begin{align}
\label{eq:pf}
{Z}(\beta) &= \Tr  e^{-\beta H}  = \int dM  [ \rho_{\cal B} (M) + \rho_{\cal F} (M) ] e^{-\beta M}
\end{align}
where $ \rho_{\cal B/F} (M) $ is the density of states in $\cal B$  and $\cal F$.  Volume independence implies 
\begin{align}
\label{eq:VI-1}
 \frac{\partial  \tilde{Z}(L)}{\partial L} = 0 
\end{align}
in the large $N$ limit.  In particular, it implies the absence of any phase transitions as the radius is reduced.

At least for $N_f < 4$, these theories are believed to be confining.  Confining large $N$ gauge theories have an infinite number of weakly-coupled stable hadronic states\cite{Witten:1979kh}.  Moreover, it is believed that confinement implies that the density of states has a stringy  `Hagedorn scaling', with the number of high-lying hadronic states increasing exponentially with mass\cite{Hagedorn:1965st,*Fubini:1969qb,*Huang:1970iq}:
\begin{align}
\label{eq:HagedornScaling}
\rho(M) \to \frac{1}{M}\left(\frac{T_H}{M}\right)^{a} e^{M/T_H} .
\end{align}
Here $T_H \sim \Lambda$ is the `Hagedorn temperature', and the values of $T_H, a$ depend on the particulars of the theory.  
Heuristically, this is motivated by the notion that at large $N$ highly excited hadrons can be described by some effective weakly-coupled string theory, and quantized relativistic strings naturally give rise to \eqref{eq:HagedornScaling}~\cite{Green:1987sp,Teper:2009uf,*Bringoltz:2005xx}.  
Hagedorn scaling for bosonic states in large $N$ theories has recently been shown directly from the assumptions of confinement and asymptotic freedom\cite{Cohen:2009wq,*Cohen:2011yx}. For QCD(Adj) the demonstration of \cite{Cohen:2009wq,*Cohen:2011yx} also implies a Hagedorn spectrum for the fermionic states, so $\rho_{\cal B},\rho_{\cal F}$ both scale as \eqref{eq:HagedornScaling}.

In a gauge theory obeying \eqref{eq:HagedornScaling}, $Z(\beta)$ diverges at $\beta^{-1} = T_H$. This implies the existence of a phase transition to a deconfined phase\cite{Thorn:1980iv} at some $T = T_d \le T_H$. Hence large $N$ volume independence cannot hold for arbitrary $\beta$ in thermal-compactified theories.  

For spatial compactifications the relevant object is $\tilde{Z}(L)$, and the story is more subtle. 
  Clearly, QCD(Adj) with $N_f=1$ is just $\mathcal{N}=1$ super-Yang-Mills theory.  Then all positive energy states are Bose-Fermi paired due to supersymmetry, and their contributions to $\tilde{Z}$ cancel, including the Hagedorn growths in  $\rho_B$ and $\rho_F$. In this case $\tilde{Z}(L)$ is the supersymmetric index $I_S$ with e.g. $I_S = N$ for $SU(N)$, and is indeed independent of $L$.   
 
Once $N_f>1$, QCD(Adj) is not supersymmetric.  If $\rho_{\cal B} - \rho_{\cal F}$ scaled as \eqref{eq:HagedornScaling} with $L=L_H$, the theory would have a phase transition.  Yet volume independence still implies \eqref{eq:VI-1} and the absence of phase transitions as a function of $L$.  In view of \eqref{eq:tpf} and \eqref{eq:HagedornScaling}, this implies the {\it equality} of all the exponentially-growing parts (not only the leading Hagedorn growth) of $\rho_{\cal B,F}$ at large $N$, which requires degeneracies between an infinite number  of bosonic and fermionic states.   What feature of QCD(Adj) could drive the equality of the Hagedorn growths in  $\cal B$  and $\cal F$?  Could a non-supersymmetric gauge theory develop emergent fermionic symmetries at large $N$, which enforce the spectral degeneracy? 

 At finite $N$ there are no-go theorems forbidding this possibility.  The Coleman-Mandula and Haag-Lopuzhansky-Sohnius theorems\cite{Coleman:1967ad,*Haag:1974qh}  imply that for theories with non-trivial S-matrices, the only allowed fermionic symmetry which acts on physical states is supersymmetry.  But hadron interactions are suppressed by powers of $1/N$, so that for $N \to \infty$ the S-matrix becomes trivial!   So these  theorems cannot be used to rule out the possibility of emergent fermionic symmetries in large $N$ gauge theories.  Nevertheless, no such emergent fermionic symmetries are known, nor is there any previously known reason to expect them.

%%%%%%%%%%%%%%%
{\it Why is QCD(Adj)  special?} 
%%%%%%%%%%%%%%%%%
There are two interrelated answers to this question, one involving the structure of the spectrum of the Hamiltonian in bosonic and fermionic Hilbert spaces, and the other concerning the one-loop analysis for a Wilson line $\Omega$ winding around the $S^1$.

First, consider $SU(N)$ Yang-Mills theory with fermions in representation ${\cal R}=\{\rm F, AS, S, Adj\}$. The first three are complex representations (fundamental and two-index anti/symmetric), whereas adjoint is a real representation.   {\it Only} fermions in the adjoint representation  endowed with PBCs can stabilize center symmetry, and hence keep the theory in the confined phase.  In particular, for all complex representations,  there is a deconfinement scale 
 $\beta_d\sim \Lambda^{-1} $ for thermal compactification with anti-PBCs, and  quantum phase transition scale 
  $L_c \sim \Lambda^{-1} $  for spatial compactification with PBCs.  

This can be viewed as a consequence of the fact that there is no crucial difference between the thermal and twisted partition functions for  ${\cal R}=\{\rm F, AS, S \}$.  In these theories the masses of color singlet states   ${\mathfrak b} \in \cal B$ scale as $m_{\mathfrak b} = O(N^0)$ whereas   ${\mathfrak f} \in {\cal F}$ scale as $m_{\mathfrak f} = \mathcal{O}(N^1)\,\,\mathrm{or}\,\, \mathcal{O}(N^2)$\cite{Witten:1979kh,Bolognesi:2006ws,*Cherman:2006iy}. Hence, in the large-$N$ limit with $L\sim \mathcal{O}(N^0)$ [or $\beta \sim \mathcal{O}(N^0)$],  the fermionic Hilbert space ${\cal F}$ is not populated, and does not contribute in any essential way to $Z$ or $\tilde{Z}$. Consequently, $\tilde{Z}(L) \approx {Z}(\beta) \approx Z_{\cal B}(\beta)$.  In  pure Yang-Mills theory, as well as large-$N$ QCD with complex representation  fermions,  
 volume independence only holds for $L>L_c \sim \Lambda^{-1}$~\cite{Kiskis:2002gr}, due to a phase transition at a $\beta = T_d^{-1}$ or $L = L_c$, and $L_c \approx T_d^{-1}$.
 
 In sharp contrast,  in  QCD(Adj), and obviously in ${\cal N}=1$ SYM,  the masses of  typical color singlet states are $m_{\mathfrak b} \sim m_{\mathfrak f}  \sim  \mathcal{O}(N^0)$.  Thus, both ${\cal B}$  and ${\cal F}$   contribute to the thermal  and twisted partition functions on the same footing, and hence we should expect no cancellation in 
$ {Z}(\beta)$  and  some degree of cancellation in $\tilde{Z}(L)$!  

Second, in QCD(Adj) with PBCs for fermions, the perturbative one-loop potential for the Wilson line  $\Omega$ is
\begin{align}
V(\Omega) =(N_f-1) \frac{2}{\pi^2 L^4} \sum_{n=1}^{\infty} \frac{1}{n^{4}} \left|\Tr \Omega^n\right|^2,
\end{align}
which is minimized at $\Omega = \eta \, {\rm diag}(1, e^{i \frac {2\pi}{N}}, \ldots,  e^{i \frac {2\pi (N-1)}{N}})$ where 
$\eta= e^{i \frac {\pi}{N}}$ for  even $N$ and one otherwise.  This is the center-symmetric vacuum\cite{Kovtun:2007py}, so at small $L$ the theory confines.
The crucial positivity of the potential is a consequence of considering the twisted partition function \eqref{eq:tpf}.   So unlike  for pure YM or ${\cal R}=\{\rm F, AS, S \}$\cite{Gross:1980br,*Unsal:2006pj}, perturbative calculations, which in this case are reliable for $N L \Lambda \ll 1$,  give no reason to expect  a center-symmetry changing phase transition (the counter-part of deconfinement in thermal case) between the 
  small-$L$ and large-$L$ regimes.    Indeed, all analytic and numerical investigations to date are consistent with the expectation that large $N$ volume independence in QCD(Adj) holds for arbitrarily small $L\sim \mathcal{O}(N^0)$
 ~\cite{Cossu:2009sq,*Bedaque:2009md,*Bringoltz:2009mi,*Bringoltz:2009kb,*Hietanen:2009ex,*Poppitz:2009fm,*Azeyanagi:2010ne,*Poppitz:2010bt,*Hietanen:2010fx,*Dorigoni:2010jv,*Catterall:2010gx,*Bringoltz:2011by,*Armoni:2011dw,*GonzalezArroyo:2012st,*Gonzalez-Arroyo:2013bta,*Gonzalez-Arroyo:2013gpa}.

{\it The clash.}
The reason to worry about the relation between Hagedorn instabilities and volume independence becomes clear if one understands the heuristic reason for the emergence of the scale $L_c \sim \Lambda^{-1}$ in e.g. pure Yang-Mills theory. 
In the confined phase
the hadronic density of states is $\mathcal{O}(N^0)$
and hadrons do not interact at large $N$.   As the temperature gets close to $T_H$, the $1/N$ suppression of interactions becomes overwhelmed by the exponential growth in the number of accessible states due to the Hagedorn scaling, and volume independence must 
break down  due to the deconfinement phase transition\cite{Cohen:2004cd,*ShifmanPrivate:2012}. 
   Our task is to see how  QCD(Adj) on $\mathbb{R}^3 \times \mathbb S^1$ with PBCs could avoid this naively inevitable breakdown. 
   
   On general grounds, we expect that $\rho_{\cal B}(M) \sim e^{L_{H}^{{\cal B}}M} $ and $\rho_{\cal F}(M) \sim e^{L_{H}^{{\cal F}}M}$, with $L_{H}^{{\cal B},{\cal F}} = c_{\cal{B},\cal{F}} \Lambda^{-1}$
where $c_{\cal{B},\cal{F}}$ are pure numbers.  However, in general there is no reason to expect that $c_{\cal{B}}=c_{\cal{F}}$.
  But if volume independence holds, as is strongly suggested by \cite{Cossu:2009sq,*Bedaque:2009md,*Bringoltz:2009mi,*Bringoltz:2009kb,*Hietanen:2009ex,*Poppitz:2009fm,*Azeyanagi:2010ne,*Poppitz:2010bt,*Hietanen:2010fx,*Dorigoni:2010jv,*Catterall:2010gx,*Bringoltz:2011by,*Armoni:2011dw,*GonzalezArroyo:2012st,*Gonzalez-Arroyo:2013bta,*Gonzalez-Arroyo:2013gpa},  the arguments above imply that the  twisted density of states 
  %$ \tilde \rho \equiv \rho_{\cal B}  - \rho_{\cal F}  $
  \begin{align}
   \tilde \rho \equiv \rho_{\cal B}  - \rho_{\cal F} 
  \end{align} 
  must obey  
$  \tilde \rho (M)  \lesssim e^{ML_H/N^p}, \;  p >0$, so that any instability associated with the growth of $\tilde{\rho}$ is suppressed at $N=\infty$ and $L_c \sim  \Lambda^{-1}/N^p$.  This is a rather  strong condition on the Hilbert space structure a theory, which we argue entails a cancellation between an infinite number of states, which in turn requires a fermionic symmetry.   

For $N_f=1$, how this can happen is obvious, as we already said above.  The masses and degeneracies of bosonic and fermionic states are indeed related by a fermionic symmetry, the  $\mathcal{N}=1$ supersymmetry, so that $\rho_{\cal B}(M) = \rho_{\cal F}(M)$, 
and $\tilde{\rho}$ does not exhibit Hagedorn growth.  The challenge is to understand the situation in the non-supersymmetric $N_f>1$ theories.

%%%%%%%%%%%%%%
{\it QM toy model.} 
%%%%%%%%%%%%%%
To develop some intuition about how the required cancellations could happen without supersymmetry, it is instructive to warm up with a toy model in quantum mechanics.  The toy model will illustrate the point of principle that high-lying contributions to $\tilde{\rho}$ can cancel even in the absence of supersymmetry.  

Consider a system comprised of one bosonic and $N_f$ fermionic harmonic oscillators, all with the same frequency $\omega$, with the Hamiltonian
\begin{align}
H = \omega \left(a^{\dag} a + \frac{1}{2} \right) 
+ \omega \sum_{i=1}^{N_f} \left(f^{\dag}_i f_i - \frac{1}{2}\right).
\end{align}
Here $a, f_i$ are bosonic/fermionic ladder operators, with the usual commutation/anticommutation  relations 
$[a,a^{\dag}]=1, \{f_i,f_i^{\dag} \}=1$. %and are meant to be caricatures of the gluons and adjoint fermions in QCD(Adj). 
 At $N_f =1$ the system becomes supersymmetric, with a single fermionic conserved charge.  However, in this free theory we can actually define $N_f$ conserved fermionic charges any $N_f \ge 1$:
\begin{align}
Q_i = a^{\dag} f_i
\end{align}
which obey $Q_i^2 = 0$.

\begin{figure}[htbp]
\begin{center}
\includegraphics[width=.48\textwidth]{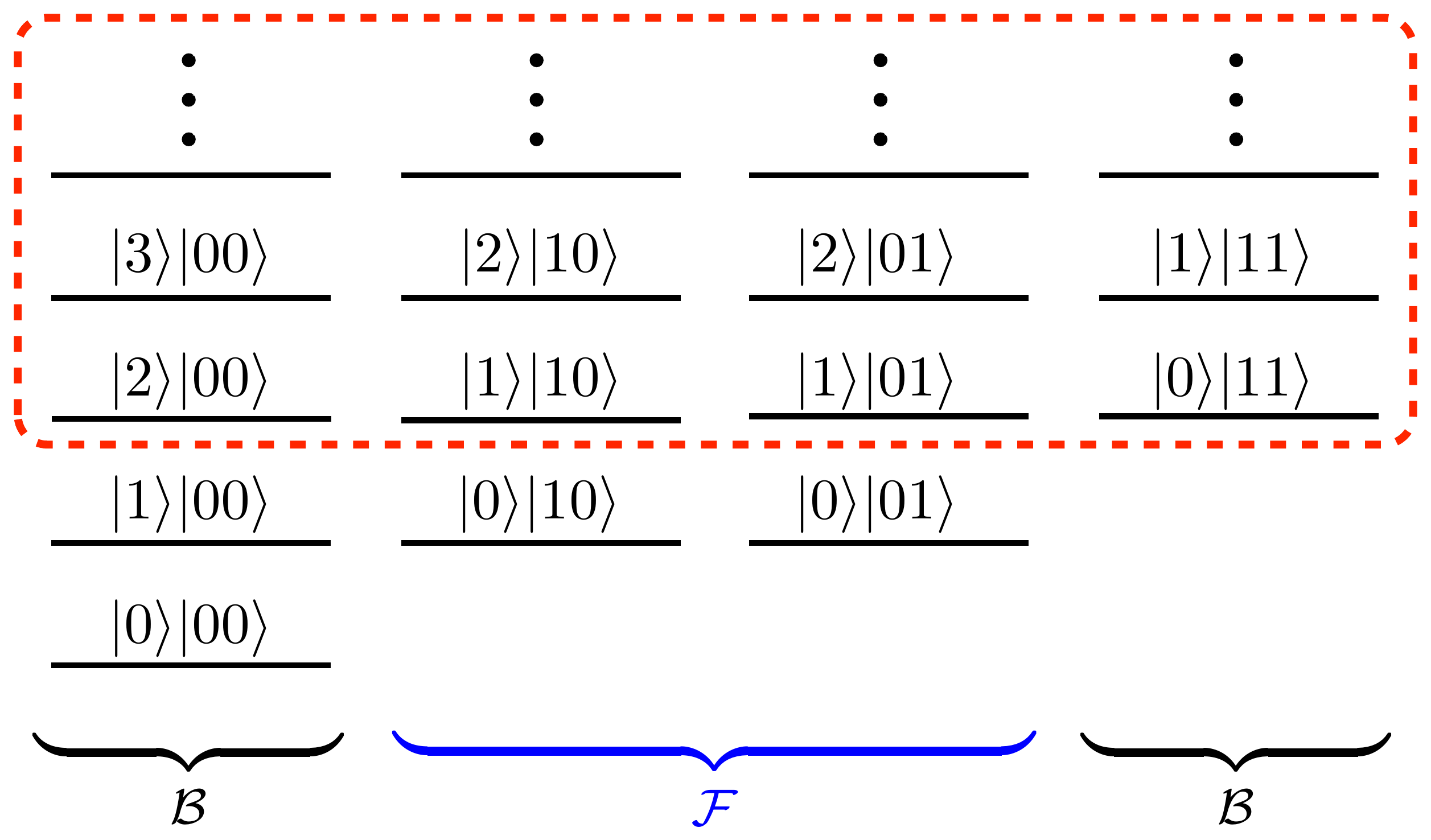}
\caption{Hilbert space  of the non-supersymmetric $N_f=2$ theory. All states in the box are paired and their contributions to $\tilde{Z}$ vanish. The low lying states saturate  $\tilde{Z}(L)$. This cancellation is also inherent to our stringy model exhibiting Hagedorn growth, where this type of cancellation occurs for each oscillator mode.}
\label{spec-1}
\end{center}
\end{figure}

This fermionic symmetry has dramatic consequences for $\tilde{Z}(L)$: the only contributions to \eqref{eq:tpf} 
come from a finite number of low-lying modes thanks to a complete cancellation of contributions from the high-lying states.  

To see how this works, let $|\psi\rangle$ be an eigenstate of the Hamiltonian $H$ with energy $E_{\psi}$ satisfying  $Q_i |\psi \rangle = 0, \forall i = 1, \ldots, N_f$.
  We can generate all the states of the system with the same energy by acting with the $Q^{\dagger}_i$'s.  In particular, consider the set of states 
 \begin{align}
 S = \{  |\psi\rangle, \; Q_i^{\dag} |\psi\rangle, \;  Q_i^{\dag} Q_j^{\dag} |\psi\rangle, \; \ldots,  \; Q_i^{\dag} \ldots Q_{i_{N_f}}^{\dag} |\psi\rangle\}
\end{align}
 Denote  the Fock states as $| n\rangle_b  | m_{1} \ldots m_{N_f} \rangle_f$. 
  If $n \geq N_f$, this procedure creates $2^{N_f}$ degenerate states at each level:   $2^{N_f-1}$ live in ${\cal B}$   and $2^{N_f-1}$ live in ${\cal F}$.   Since these states contribute to the twisted partition function with opposite signs, their net effect vanishes. 
   For $n < N_f$, some combinations of $Q_{i_1}^{\dag}\ldots Q_{i_k}^{\dag} $'s annihilate $\psi$, and the number of states living in  ${\cal B}$ and ${\cal F}$ are unequal.  This is depicted for $N_f=2$ in Fig.~\ref{spec-1}. Therefore, the twisted partition function is saturated by low 
   states  with $n < N_f$, and
\begin{align}
\label{eq:QMTwistedZ}
\tilde{Z}(L)&= \tilde{Z}_{n< N_f} +\tilde{Z}_{n \ge N_f} =\tilde{Z}_{n< N_f}    \cr
 & = \left[2\sinh\left(\frac{\omega L}{2}\right)\right]^{(1-N_f)/2}
\end{align}  
which can derived using combinatorial identities. Note that here $\tilde{Z}$ gets contributions from finite-energy states, and not just from zero modes as in the SUSY case.

Of course, the second line of \eqref{eq:QMTwistedZ} follows immediately if one evaluates the path integral for a single real scalar field and $N_f$ fermions living on a circle of length $L$, all with a common mass $\omega$, and periodic boundary conditions for the fermions.   By considering $ \Tr (-1)^F e^{-L H}$ explicitly as a graded sum over Hilbert space,  however, we have seen that this innocuous-seeming result is actually the consequence of a  fermionic symmetry  which leads to a particular pattern of cancellations.

%%%%%%%%%%%%%
{\it  Stringy Toy Model with Hagedorn growth.}
%%%%%%%%%%%%%
We now consider a `stringy' toy model with $\rho$ scaling as \eqref{eq:HagedornScaling}.  A confining large $N$ theory like QCD(Adj) is expected to have bosonic and fermionic `glueball' spectra that asymptotically come to lie on Regge trajectories, see e.g. \cite{Meyer:2004gx}.  Taking this as an inspiration, our toy model is defined to have the spectrum $M^2 = N/\alpha'$, where 
\begin{equation}
\label{eq:StringyHamiltonian}
N= \sum_{n\in\mathbb{N}} n\,a_n^\dagger a_n+  \sum_{i=1}^{N_f} \sum_{n\in\mathbb{N}} n\,f_{i\,n}^\dagger f_{i\,n}\,,
\end{equation}
$a_n^\dagger, a_n$ are bosonic creation/annihilation operators, and $f_{i\,n}^\dagger,f_{i\,n}$ are their fermionic counterparts. This model is meant to illustrate, as point of principle, how the Hagedorn instability can be avoided in a non-supersymmetric theory.  

The asymptotic behavior of $\rho(M)$ is determined by the number of states with $M = \sqrt{n/\alpha'}$.   To evaluate this, we count how many states there are with $M^2 = n/\alpha'$ by considering  the combinatorial generating function 
\begin{equation}
\label{eq:ThermalGeneratingFunction}
\Tr q^N = \prod_{n=1}^{\infty} \frac{(1+q^n)^{N_f}}{1-q^n} =\sum_{n=0}^\infty d(n)\,q^n\,,
\end{equation}
where $d(n)$ are the desired degeneracy factors.  Note that when $N_f=0$, for example,  $d(n)$ counts  the number of integer partitions of $n$.
Standard methods\cite{Green:1987sp} yield 
\begin{equation}
\label{eq:ThermalStringyScaling}
d(n)\sim \, \exp \left(\sqrt{  2 \pi^2  (1+N_f/2)  n /3  } \right),\quad n\gg 1,
\end{equation}
and $\rho(M)$ follows \eqref{eq:HagedornScaling} with 
$T_H^{-1} =  \sqrt{\alpha' 2 \pi^2  (1+N_f/2) /3 }$.  
As expected this stringy model has a Hagedorn instability which makes $Z(\beta)$  diverge for $T>T_H$.

We notice that even when $N_f>1$ and the theory has a different number of bosonic and fermionic modes, we can still define $N_f$ fermionic operators
\begin{equation}
\label{eq:StringySuperSUSYcharges}
Q_i = \sum_{n\in\mathbb{N}} \sqrt{n} \, a^\dagger_n f_{i\,n}    ,
\end{equation}
satisfying $\{Q_i,Q_j\}=0$. 
These charges commute with the mass square operator $[Q_i,M^2]=0$, without any violation of Coleman-Mandula-type theorems since the theory is free. The $Q-$cohomology of these operators generates the full Hilbert space, just as in the previous toy model. That is, we can obtain all of the eigenstates of $M^2$ with eigenvalue $n/\alpha'$ by starting with all $Q-$closed states $\vert\psi \rangle$: $Q_i \vert\psi \rangle=0\,\forall\,i=1,...,N_f$ s.t. $\alpha' M^2 \vert\psi \rangle= n \vert\psi \rangle$.  Then acting with $Q_i^\dagger$ we obtain all the other degenerate states: $\{ Q_{i}^\dagger \vert\psi \rangle,...,Q_{i_1}^\dagger ...Q_{i_{N_f}}^\dagger\vert\psi \rangle \}$.
When $N_f=1$ this symmetry is just supersymmetry, and states with $M^2>0$ are all Bose-Fermi paired.  However, for $N_f > 1$, the number of bosonic and fermionic states is not generally the same.  Does this mean that in this model $\tilde{Z}(L)$ has a Hagedorn instability for $N_f \geq 2$, similarly to $Z(\beta)$? 
 
Perhaps surprisingly, thanks to the fermionic symmetry \eqref{eq:StringySuperSUSYcharges}, the answer is no.
To understand the behavior of $\tilde{\rho}(M)$ for high $M$, we now consider
\begin{equation}
\Tr \left[ (-1)^F q^{N}\right]= \prod_{n=1}^\infty {(1-q^n)^{(N_f-1)}}=\sum_{n=0}^\infty c(n) q^n
\end{equation}
It turns out that $c(n)$ does not grow exponentially once $N_f \ge 1$.  To get a feeling for how this happens, consider the simple case of $N_f=2$, where 
\begin{align}
\label{eq:TwistedGeneratingFunction}
&\Tr \left[ (-1)^F q^{N}\right] =1+\sum_{n=1}^\infty\left[ (-1)^n  q^{\frac{3n^2-n}{2}}+ (-1)^n  q^{\frac{3n^2+n}{2}} \right]\nonumber \\
&=1-q-q^2+q^5+q^7 -q^{12} - q^{15} + q^{22} + \ldots  .
\end{align}
The cancellation between fermions and bosons with the same energy is striking: between the first $200$ energy levels only $23$ have $\tilde{\rho}(M)\neq 0$ and thanks to Euler's pentagonal number theorem, the difference in populations is either $0$ or $\pm 1$.   Indeed, as with \eqref{eq:ThermalGeneratingFunction}, the twisted generating function \eqref{eq:TwistedGeneratingFunction} counts the partitions of the integers, but in a ``graded" way. The partitions with an even number of terms are counted positively, whereas partitions with an odd number are counted negatively.  
The levels at which the $(-1)^n$-mismatch occurs  are called generalized pentagonal numbers, $p^{\pm}_n = (3n^2 \pm n)/2$. 

Despite the fact that the twisted partition function is not an index,  and in fact receives contribution from infinitely many states, with Hagedorn growths in $\cal{B}$ and $\cal{F}$, the contributions from ${\cal B}$ and ${\cal F}$  cancel almost perfectly  except for the levels $p^{\pm}_n$, by $(-1)^n$.  This is so tame that it does not lead to a Hagedorn instability. This generalizes to $N_f>2$, where for  $n\gg 1$ one can show the bound $|c_n| < n^{N_f-2}$, and $\tilde{\rho}$ does not grow exponentially.  

%%%%%%%%%%%%%%%%%%%%%
{\it Conclusions.} 
%%%%%%%%%%%%%%%%%%%%%
We have explained the apparent tension between a Hagedorn density of states and large $N$ volume independence in circle-compactified QCD(Adj).  If, as expected, QCD(Adj) has both of these properties, there is a striking implication: at large $N$ we expect the emergence of a fermionic symmetry for $N_f > 1$, leading to cancellations which evade the Hagedorn instability.  We showed an explicit example of how this can happen in a toy model.    An emergent symmetry of this sort may have phenomenological implications due to the large $N$ `orientifold equivalence' connecting QCD(Adj) to QCD(AS) at large $L$, with the latter being a natural large $N$ limit of real QCD\cite{Armoni:2003gp,*Armoni:2004uu}.  Clearly, there are many directions for future analytic and numerical work. For instance, lattice calculations can test the idea by looking for spectral degeneracies when $N_f > 1$. Analytically, it would be fascinating to better understand the nature of this new kind of emergent symmetry, and explore its implications. For example, at $N_f = 1$ the vacuum energy is zero due to supersymmetry, but this will not be the case for $N_f >1$. But the $N_f>1$ emergent fermionic symmetry may still lead to highly nontrivial cancellations, which seems like an interesting direction for future work. 

{\it Node Added.}  After the publication of this work, we became aware of the relevance to our discussion of the idea of misaligned supersymmetry\cite{Dienes:1994np,*Dienes:1994jt,*Dienes:1995pm}, which was mostly discussed before in the context of non-supersymmetric string theories.  Refs.~\cite{Dienes:1994np,*Dienes:1994jt,*Dienes:1995pm} point out that demanding the absence of tachyons for consistency of such theories, implemented via the requirement of modular invariance, implies that all Hagedorn-growing contributions to the density of states from bosons and fermions cancel in the twisted partition function.  Hence it is plausible that $N_f>1$ QCD(Adj) at large $N$ may be the first known example of a field theory exhibiting misaligned supersymmetry. 

%%%%%%%%%%%%%%%%%%%%
{\it Acknowledgements.}
%%%%%%%%%%%%%%%%%%%%
We are indebted to Misha Shifman, whose probing questions about these issues served as the inspiration for this project, and for useful comments on the paper.  We also thank Ali Altu\u{g}, Adi Armoni, Tom Cohen, Keith Dienes, Gerald Dunne, Michael Green, Biagio Lucini, Robert Myers, and Sungjay Lee for very useful and stimulating discussions.   We acknowledge support from DOE grants  DE-FG-88ER40388 (G.~B.),
 FG02-94ER40823 (A.~C.),  DE-FG02-12ER41806 (M.~\"U.) and
European Research Council Advanced Grant No. 247252 ``Properties and Applications of the Gauge/Gravity Correspondence'' (D.~D.).

\bibliographystyle{apsrev4-1}
\bibliography{super_susy} 

\end{document}